\documentclass[conference]{IEEEtran}
\IEEEoverridecommandlockouts
\usepackage{cite}
\usepackage{amsmath,amssymb,amsfonts}
\usepackage{algorithmic}
\usepackage{graphicx}
\usepackage{textcomp}
\usepackage{xcolor}
\usepackage{graphicx}
\usepackage{subfigure}
\usepackage{multirow}
\usepackage{booktabs}
\usepackage{caption}
\usepackage{booktabs}
\usepackage{hyperref}
\usepackage[numbers,sort&compress]{natbib}
\usepackage{url}

\def\BibTeX{{\rm B\kern-.05em{\sc i\kern-.025em b}\kern-.08em
    T\kern-.1667em\lower.7ex\hbox{E}\kern-.125emX}}
\begin{document}

\title{NF-TrackLLM: Joint Prediction of UAV Trajectory and Near-Field Beam for LAE XL-MIMO Systems \\
\thanks{Corresponding authors: Y. Han, S. Jin.}
}

\author{Qianfan Lu$^\ast$, Mengyuan Li$^\ast$, Jiachen Tian$^\ast$, Yu Han$^\ast$, Xiao Li$^\ast$, and Shi Jin$^\ast$\\
$^\ast$School of Information Science and Engineering, Southeast University, Nanjing 210096, China\\
Email: \{qianfan\_lu, mengyuan\_li, tianjiachen, hanyu, li\_xiao, jinshi\}@seu.edu.cn
}

\maketitle

\begin{abstract}
User localization and beam management are tightly linked in extremely large-scale multiple-input multiple-output (XL-MIMO) systems, especially in dense low-altitude economy (LAE) scenarios.
However, the near-field propagation in XL-MIMO introduces strong distance sensitivity and complex spatial coupling, which makes joint trajectory and beam prediction challenging. 
Meanwhile, large language models (LLMs) have attracted attention in physical-layer transmission for modeling long-range dependencies.
In this paper, we propose NF-TrackLLM, a multi-modal semantic-aware framework for near-field unmanned
aerial vehicles (UAVs) positioning and beam prediction in XL-MIMO systems.
By incorporating visual and LiDAR sensing into a Sionna-based channel generation pipeline, environmental semantics and GPS are utilized to guide trajectory and beam prediction. Built upon the aligned multi-modal representation, a GPT-2-based spatiotemporal reasoning backbone, and a cascaded prediction strategy are employed, where future trajectories are first inferred and then used to guide beam prediction as geometric priors. 
Simulation results demonstrate that NF-TrackLLM achieves accurate beam prediction and reliable UAV trajectory tracking in dense urban low-altitude scenarios.

\end{abstract}


\section{Introduction}
The sixth-generation (6G) wireless communication system is expected to support ultra-high data rates, ultra-low latency, and massive connectivity, which has stimulated the deployment of extremely large-scale multiple-input multiple-output (XL-MIMO) \cite{XL-MIMO_Systems}.
User positioning and beam management are identified as exemplary AI use cases prioritized by 3GPP RAN1 \cite{3gpp.38.843}, highlighting their importance in future networks.
However, as the antenna array size increases, near-field effect becomes prevalent \cite{near-field}\cite{XLMIMO_nearfield}, making accurate user localization and beam management more challenging, especially for highly mobile users such as unmanned aerial vehicles (UAVs).
Meanwhile, as 6G evolves toward intelligence-enabled paradigms, reconstructing the physical-layer transmission mechanisms with large language models (LLMs) has emerged as a promising direction, owing to their capability in capturing long-range dependencies. 
Despite recent advances in beam management, joint user positioning and beam prediction in XL-MIMO near-field scenarios, especially under long-horizon temporal evolution, remains insufficiently studied.

Among existing learning-based beam prediction methods, some approaches incorporate multi-modal information to improve prediction accuracy, as different modalities can capture complementary aspects of the environment and user dynamics.
For instance, GPS and images collected at the base station (BS) have been used for beam prediction \cite{1-1}, while LiDAR-based point clouds are introduced to enhance robustness under unfavorable lighting conditions \cite{1-3}. More recent works combine vision, LiDAR, radar sensing, and GPS to assist beam prediction \cite{2-1,2-3}.
Beyond multi-modal approaches, some studies address beam prediction by modeling temporal correlations in historical beam sequences \cite{4-4}.
While effective in the far-field, these approaches overlook the spherical wavefront propagation in XL-MIMO. Unlike far-field steering, near-field communications rely on beam focusing, which necessitates precise matching in both angular and distance dimensions. 
Neglecting the distance dimension renders existing datasets and algorithms insufficient for 3D angle-distance beam prediction.
Moreover, due to the fast-varying nature of channels, long-term prediction relying solely on past channels is unreliable. In contrast, environmental semantics, such as building structures and road layouts, change slowly over time. Therefore, how to effectively incorporate such information into trajectory and near-field beam prediction remains a critical problem.

In this paper, we propose NF-TrackLLM, a multi-modal semantic-aware framework for near-field UAV localization and beam prediction in low-altitude economy (LAE) XL-MIMO systems. We extend a Sionna-based LAE XL-MIMO dataset\footnote{Dataset: \url{https://huggingface.co/datasets/lmyxxn/MultimodalNF}; Generator: \url{https://github.com/Lmyxxn/Multimodal-NF}}\cite{dataset} by incorporating visual and LiDAR modules.
To address the distance sensitivity of near-field beams and the limited reliability of historical information, NF-TrackLLM exploits environmental semantics via multi-modal feature alignment and fusion of images, LiDAR, and UAV GPS, improving robustness under NLoS conditions.
A GPT-2-based \cite{gpt2} spatiotemporal reasoning backbone is then employed.
Furthermore, a cascaded prediction head is developed, in which future UAV trajectories are first inferred and subsequently used as priors to predict beams.
Simulation results demonstrate that NF-TrackLLM achieves accurate beam prediction and UAV trajectory tracking in challenging urban LAE scenarios.

\enlargethispage{1\baselineskip}
\section{System Model}

\begin{figure}[!t]
\centering
\includegraphics[height=4.3cm,width=8cm]{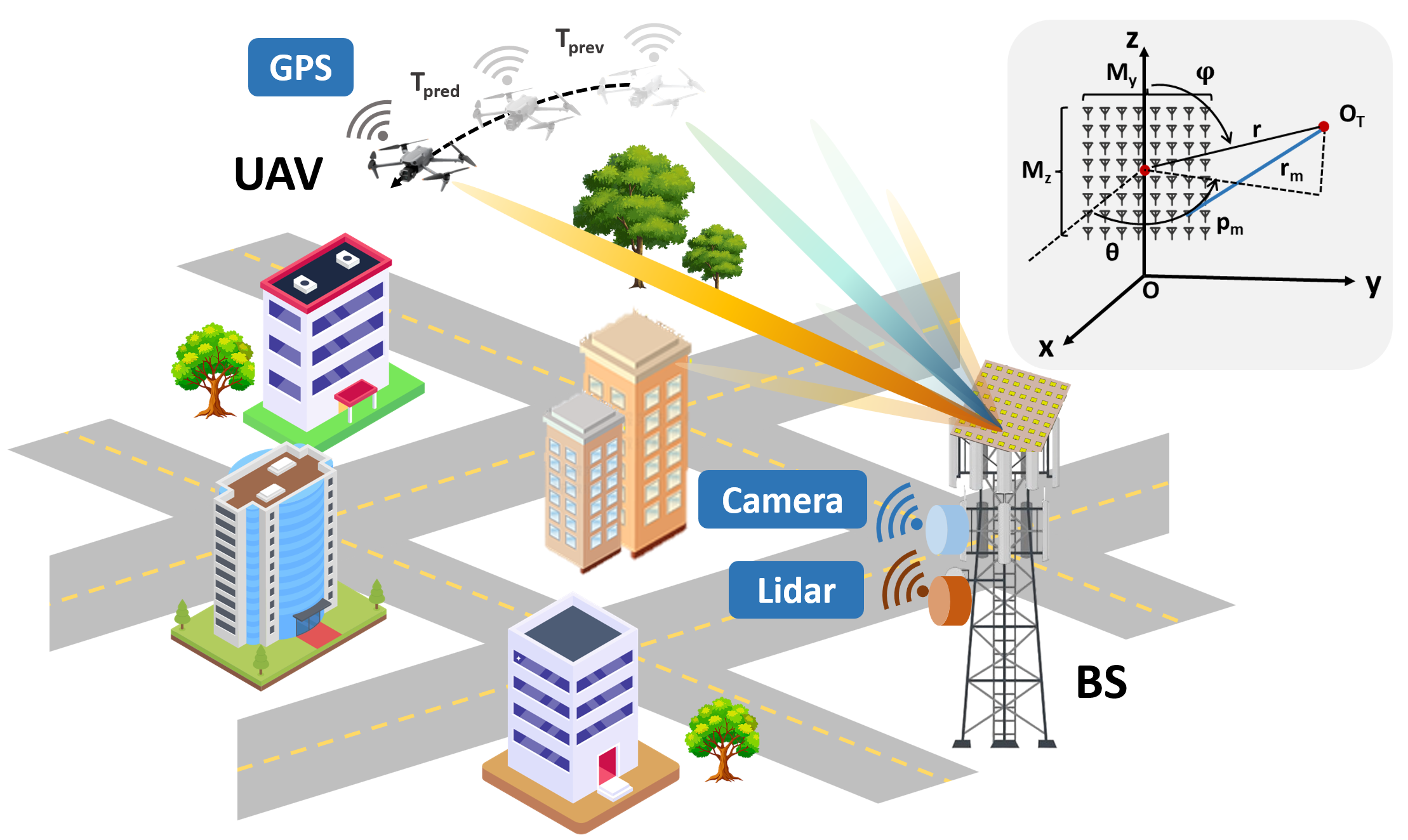}
\caption{Illustration of the XL-MIMO system model: The BS is equipped with a camera and a LiDAR, while the UAV transmits GPS coordinates to the BS.}
\label{1}
\end{figure}

We consider a single-cell XL-MIMO system consisting of a BS and a mobile UAV operating within the near-field region in a low-altitude scenario, as illustrated in Fig.~\ref{1}. The BS employs a uniform planar array (UPA) with $M_y \times M_z$ antennas, where $M_y$ and $M_z$ denote the number of antennas along the $y$- and $z$-axes, respectively. The carrier frequency is denoted as $f_c$ with wavelength $\lambda$. The UAV is equipped with a single antenna and follows a time-varying three-dimensional trajectory.
The antenna spacings $d_y$ and $d_z$ are both set to $0.5\lambda$. 
Both line-of-sight (LoS) and non-line-of-sight (NLoS) paths are considered.
The time-varying near-field uplink channel of the $m$-th antenna at time $t$ can be expressed as
\begin{equation}
    h_m(t)=\sum_{l=1}^{L_m(t)}
    g_{l,m}(t)e^{-j k r_{l,m}(t)},
\end{equation}
where $k = 2\pi /\lambda $, $L_m(t)$ denotes the number of paths, $r_{l,m}(t)$ represents the distance between the $l$-th scatterer and the $m$-th antenna, and $g_{l,m}(t)$ is the corresponding complex path gain.

Assuming all-1 uplink pilots are sent by the UAV, the received signal after beamforming with a combining vector $\mathbf{w}$ can be expressed as
\begin{equation}
    \mathbf{y} = \sqrt{P_r}\,\mathbf{w}^{\mathrm{H}}\mathbf{h} + \mathbf{n},
\end{equation}
where $P_r$ denotes the uplink transmit power and $\mathbf{n}$ is the additive Gaussian complex noise. The optimal beam selection problem can then be formulated as
\begin{equation}
    \mathbf{w}^\star = \arg\max_{\mathbf{w}\in\mathcal{W}}
    \log_2\!\left( 1 + \frac{\left|\mathbf{w}^{\mathrm{H}}\mathbf{h}\right|^2}{\sigma^2}\right),
\end{equation}
where $\sigma^2$ denotes the noise power at the receiver, and $\mathcal{W}$ is the near-field beamforming codebook, constructed by jointly sampling the angular and distance domains:
\begin{equation}
    \mathcal{W} =
    \big[
    \mathbf{b}(\theta_1,\phi_1,r_1),
    \ldots,
    \mathbf{b}(\theta_1,\phi_1,r_S),
    \ldots,
    \mathbf{b}(\theta_N,\phi_N,r_S)
    \big].
\end{equation}
Here, $\{\theta_i,\phi_i\}_{i=1}^{N}$ denote the azimuth and elevation angular codeword pairs, where $N$ is the number of candidate angular beam codewords. $r_s$ represents the $s$-th candidate distance and $S$ denotes the number of candidate distance codewords.
This joint angle-distance sampling results in a three-dimensional partitioning of the near-field space.
$\mathbf{b}(\theta_i,\phi_i,r_s)$ is the steering vector corresponding to the point $(\theta_i,\phi_i,r_s)$, which can be written as
\begin{equation}
    \mathbf{b}(\theta,\phi,r) =
    \big[
    e^{-jkr_0},
    \ldots,
    e^{-jkr_{M-1}}
    \big]^{\mathrm{T}},
\end{equation}
where $M = M_y M_z$ represents the total number of antennas in the UPA, and $r_m$ denotes the distance from the focal point $(\theta,\phi,r)$ to the $m$-th antenna.

\section{NF-TrackLLM Framwork}

\begin{figure*}[t]
\centering
\includegraphics[height=5.8cm,width=16.8cm]{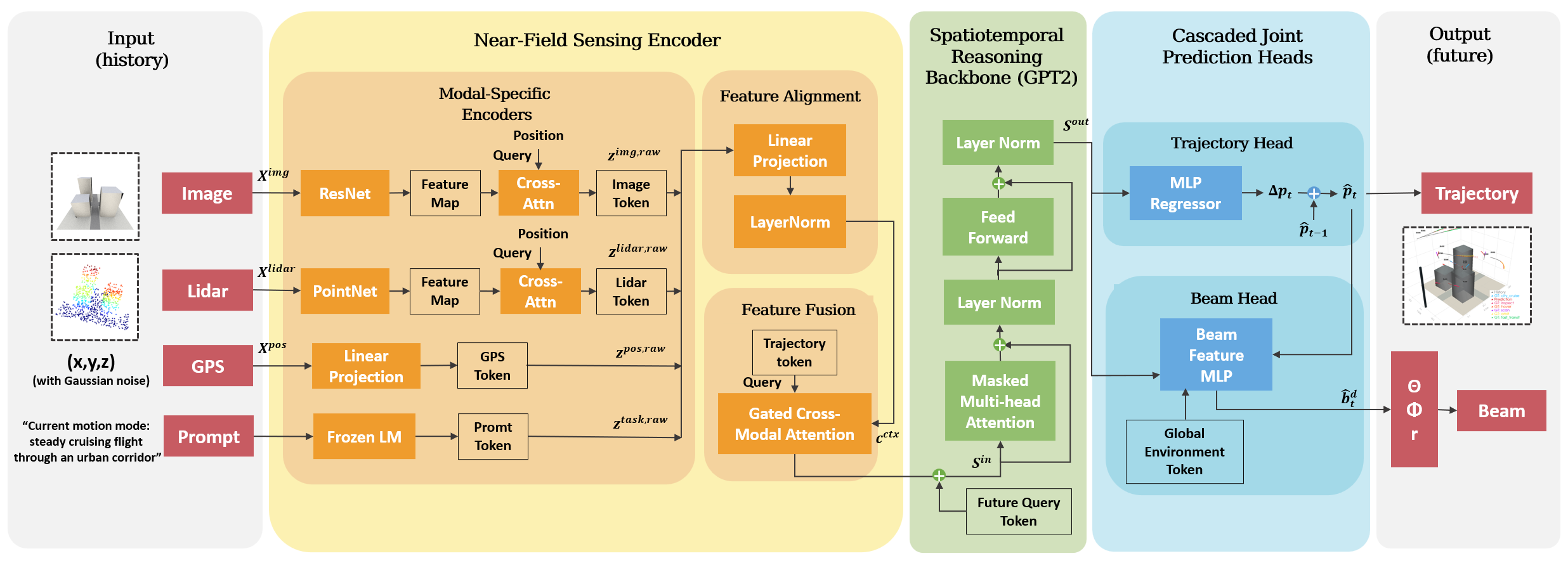}
\caption{Architecture of the NF-TrackLLM framework. Multi-modal inputs, including images, LiDAR, GPS, and prompts, are first encoded and fused into unified tokens. A GPT-2-based backbone performs spatiotemporal reasoning, followed by a cascaded trajectory-first, beam-following prediction head to infer future UAV trajectories and near-field beam indices.}
\label{2}
\end{figure*}

In this section, we propose NF-TrackLLM, which utilizes multi-modal spatiotemporal reasoning to predict near-field trajectories and beams for the future $T_{\mathrm{pred}}$ steps based on $T_{\mathrm{prev}}$ historical observations. As illustrated in Fig.~\ref{2}, NF-TrackLLM consists of three key components: a near-field sensing encoder, a spatiotemporal reasoning backbone, and cascaded joint prediction heads.

\subsection{Near-Field Sensing Encoder}

To mitigate error accumulation under long-term prediction and NLoS conditions, multi-modal near-field sensing is introduced as a stable source of environmental context. The system leverages a multi-modal input stream over $T_{\mathrm{prev}}$ historical time slots, consisting of BS-view RGB images $\mathbf{X}^{\mathrm{img}} \in \mathbb{R}^{T_{\mathrm{prev}} \times H \times W \times 3}$, LiDAR point clouds $\mathbf{X}^{\mathrm{lidar}} \in \mathbb{R}^{T_{\mathrm{prev}} \times P \times 3}$, and simulated GPS coordinates $\mathbf{X}^{\mathrm{pos}} \in \mathbb{R}^{T_{\mathrm{prev}} \times 3}$ (with Gaussian noise). $H, W$ and $P$ represent the image dimensions and the number of LiDAR points, respectively. The designed near-field sensing encoder is composed of the following modules:

\textbf{1) Modal-Specific Encoding:}
Each sensing modality is first processed by an independent encoder to extract modality-specific features. 
For visual sensing, a pre-trained ResNet \cite{resnet} is applied to the image $\mathbf{X}^{\mathrm{img}}_t$ to obtain spatial feature maps $\mathcal{E}_{\mathrm{img}}(\mathbf{X}^{\mathrm{img}}_t)$.
A position-conditioned attention mechanism is then employed using the position $\mathbf{p}_t \in \mathbb{R}^3$, i.e., the $t$-th element of $\mathbf{X}^{\mathrm{pos}}$, as query:
\begin{equation}
\mathbf{z}^{\mathrm{img},\mathrm{raw}}_t
=
\mathrm{Attn}\!\Big(
\mathcal{E}_{\mathrm{pos}}(\mathbf{p}_t),\,
\mathcal{E}_{\mathrm{img}}(\mathbf{X}^{\mathrm{img}}_t)
\Big),
\end{equation}
where $\mathcal{E}_{\mathrm{pos}}(\cdot)$ denotes a multi-layer perceptron (MLP)-based position encoder and $\mathrm{Attn}(\cdot)$ represents a multi-head attention operation.
For geometric sensing, LiDAR $\mathbf{X}^{\mathrm{lidar}}_t$ are processed by a PointNet-based \cite{pointnet} encoder to extract point-wise features $\mathcal{E}_{\mathrm{lidar}}(\mathbf{X}^{\mathrm{lidar}}_t)$.
Similarly, a position-guided attention mechanism is applied to aggregate these features:
\begin{equation}
\mathbf{z}^{\mathrm{lidar},\mathrm{raw}}_t
=
\mathrm{Attn}\!\Big(
\mathcal{E}_{\mathrm{pos}}(\mathbf{p}_t),\,
\mathcal{E}_{\mathrm{lidar}}(\mathbf{X}^{\mathrm{lidar}}_t)
\Big),
\end{equation}
which enables the model to selectively focus on environmental information surrounding the position of the UAV.
The UAV trajectory is encoded from the GPS position sequence:
\begin{equation}
\mathbf{z}^{\mathrm{pos},\mathrm{raw}}_t
=
\mathbf{W}_{\mathrm{pos}} \mathbf{p}_t,
\end{equation}
where $\mathbf{W}_{\mathrm{pos}}$ is a learnable linear projection matrix.

In addition, task-level semantic prompts are introduced to describe high-level motion patterns or environmental conditions.
For example, a prompt like “A low-altitude UAV operating in a Near-Field XL-MIMO communication system performing joint trajectory and beam prediction tasks. Current motion mode: steady cruising flight through an urban corridor” encodes prior knowledge.
Each prompt is encoded by a frozen language model to produce a semantic embedding $\mathbf{z}^{\mathrm{task},\mathrm{raw}}_t$.
A prompt set consisting of 10 predefined descriptions is constructed, and the appropriate prompt is selected based on the UAV flight mode and environmental scenario.

\textbf{2) Heterogeneous Feature Alignment:}
Since the modality-specific features are extracted by heterogeneous encoders and thus reside in different representation spaces, a feature alignment module is introduced to project all modality tokens into a shared embedding space.
For each modality $u \in \{\mathrm{img}, \mathrm{lidar}, \mathrm{pos}, \mathrm{task}\}$, the corresponding token $\mathbf{z}^{(u),\mathrm{raw}}_t$ at time slot $t$ is linearly projected and normalized as
\begin{equation}
\mathbf{z}^{(u)}_t
=
\mathrm{LN}\!\left(
\mathbf{W}^{(u)} \mathbf{z}^{(u),\mathrm{raw}}_t
\right),
\end{equation}
where $\mathbf{W}^{(u)}$ denotes the learnable projection matrix for modality $u$, and $\mathrm{LN}(\cdot)$ represents layer normalization.
This alignment step ensures that heterogeneous sensing data and semantic prompts are embedded into a unified space.

\textbf{3) Multi-Modal Feature Fusion:}
After alignment, the modality-specific tokens are concatenated along the modality dimension to form a set of environmental context tokens:
$\mathbf{c}^{\mathrm{ctx}}_t
=
\left\{
\mathbf{z}^{\mathrm{img}}_t,\;
\mathbf{z}^{\mathrm{lidar}}_t,\;
\mathbf{z}^{\mathrm{pos}}_t,\;
\mathbf{z}^{\mathrm{task}}_t
\right\}.$
Let $\mathbf{c}^{\mathrm{traj}}_t$ denote the trajectory token obtained from the position embedding $\mathbf{z}^{\mathrm{pos}}_t$ after projection into the shared embedding space.
To inject environmental semantics into motion modeling, a gated cross-modal attention module $\mathcal{F}_{\mathrm{cross}}(\cdot)$ is applied, taking $\mathbf{c}^{\mathrm{traj}}_t$ as the query and $\mathbf{c}^{\mathrm{ctx}}_t$ as key-value pairs:
\begin{equation}
\mathbf{c}_t
=
\mathcal{F}_{\mathrm{cross}}
\!\left(
\mathbf{c}^{\mathrm{traj}}_t,\,
\mathbf{c}^{\mathrm{ctx}}_t
\right).
\end{equation}
The cross-attention operation adaptively aggregates heterogeneous sensing and semantic information according to their relevance to the current trajectory state, followed by gated residual fusion and feed-forward refinement.
The resulting fused token $\mathbf{c}_t$ jointly encodes UAV motion dynamics and near-field environmental semantics, and serves as the trajectory representation for subsequent temporal reasoning.

\subsection{Spatiotemporal Reasoning Backbone}
To capture long-range temporal dependencies, GPT-2 is adopted as the spatiotemporal reasoning backbone.
The input sequence is constructed by concatenating the fused historical trajectory tokens and a set of learnable future query tokens:
\begin{equation}
\mathbf{S}_{\mathrm{in}}
=\Big[
\mathbf{c}_{1:T_{\mathrm{prev}}},
\;
\mathbf{q}_{1:T_{\mathrm{pred}}}
\Big],
\end{equation}
where $\mathbf{c}_{1:T_{\mathrm{prev}}} = [\mathbf{c}_1, \ldots, \mathbf{c}_{T_{\mathrm{prev}}}]$ denotes the cross-modally enhanced trajectory tokens, and $\mathbf{q}_t$ are learnable query tokens 
that serve as placeholders for unobserved future time slots.
After temporal embedding, the input sequence is processed by GPT-2 to produce contextualized hidden states:
\begin{equation}
\begin{gathered}
\mathbf{S}_{\mathrm{out}}
=
\mathcal{G}_{\mathrm{GPT}}\!\left(\mathbf{S}_{\mathrm{in}}\right)
=
\left[
\mathbf{s}_1,\ldots,
\mathbf{s}_{T_{\mathrm{prev}}+T_{\mathrm{pred}}}
\right], \\
\mathbf{S}^{\mathrm{pred}}
=
\left[
\mathbf{s}_{T_{\mathrm{prev}}+1},
\ldots,
\mathbf{s}_{T_{\mathrm{prev}}+T_{\mathrm{pred}}}
\right].
\end{gathered}
\end{equation}

\subsection{Cascaded Joint Prediction Heads}
A cascaded prediction head is designed to perform
trajectory-first, beam-following inference. Based on the future hidden states $\mathbf{s}_t \in \mathbf{S}^{\mathrm{pred}}$, incremental motion offsets $\Delta \hat{\mathbf{p}}_t$ are first predicted and accumulated to obtain the future trajectory $\hat{\mathbf{p}}_t$:
\begin{equation}
\Delta \hat{\mathbf{p}}_t
=
\mathcal{F}_{\mathrm{traj}}(\mathbf{s}_t),
\qquad
\hat{\mathbf{p}}_t
=
\hat{\mathbf{p}}_{t-1} + \Delta \hat{\mathbf{p}}_t ,
\end{equation}
where $\mathcal{F}_{\mathrm{traj}}(\cdot)$ denotes a trajectory prediction head implemented as a MLP.
The predicted positions are then injected into beam prediction by concatenating trajectory features, backbone representations, and global environmental context, which are jointly mapped to the beam-related representation by the beam prediction head $\mathcal{F}_{\mathrm{beam}}(\cdot)$:
\begin{equation}
\mathbf{u}_t
=
\mathcal{F}_{\mathrm{beam}}
\!\left(
\Big[
\mathbf{s}_t
\;\|\;
\mathcal{E}_{\mathrm{pos}}(\hat{\mathbf{p}}_t)
\;\|\;
\bar{\mathbf{c}}^{\mathrm{env}}
\Big]
\right),
\end{equation}
where $\bar{\mathbf{c}}^{\mathrm{env}}$ denotes the global environmental context obtained by averaging the aligned multi-modal context tokens. Finally, three parallel classification heads are applied to predict azimuth, elevation, and distance beam indices $\hat{\mathbf{b}}^{(d)}_t$:
\begin{equation}
\hat{\mathbf{b}}^{(d)}_t
=
\mathcal{F}_{d}(\mathbf{u}_t),
\quad
d \in \{\theta, \phi, r\}.
\end{equation}
This decoupled design is motivated by the intrinsic structure of near-field beam codebooks.
Directly predicting a single beam index over such a large and fragmented codebook is impractical and lacks scalability. By decoupling the beam index into three components, the proposed design reduces classification complexity and enables more stable prediction.

\subsection{Physics-Aligned Loss}
We design a physics-aligned loss function that supervises trajectory regression and beam selection.
The mean absolute error (MAE) loss is adopted for trajectory supervision:
\begin{equation}
\mathcal{L}_{\mathrm{traj}}
=
\frac{1}{B T_{\mathrm{pred}}}
\sum_{b=1}^{B}
\sum_{t=1}^{T_{\mathrm{pred}}}
\left\|
\hat{\mathbf{p}}_{b,t}
-
\mathbf{p}_{b,t}
\right\|_1 ,
\end{equation}
where $B$ denotes the batch size.
To provide tolerant supervision under near-field propagation, we adopt a soft Top-$K$ beam loss based on the Kullback--Leibler (KL) divergence.
For each future time slot $t$ and beam dimension $d \in \{\theta, \phi, r\}$, a soft target probability $p^{(d)}_{t,i}$ is assigned to each beam candidate $i$ based on the Top-$K$ ground-truth beams, where higher-ranked beams receive larger decay-weighted probabilities.
Let $\mathbf{l}^{(d)}_t$ denote the predicted beam logits for dimension $d$ at time slot $t$. The beam loss for each dimension is defined as
\begin{equation}
\mathcal{L}^{(d)}
=
\frac{1}{T_{\mathrm{pred}}}
\sum_{t=1}^{T_{\mathrm{pred}}}
\sum_{i=1}^{N_d}
p^{(d)}_{t,i}
\log
\frac{
p^{(d)}_{t,i}
}{
\mathrm{Softmax}\!\left(\mathbf{l}^{(d)}_t\right)_i
},
\end{equation}
where $N_d$ denotes the number of beam candidates for dimension $d$.
The overall beam loss is given by $\mathcal{L}_{\mathrm{beam}} = \mathcal{L}^{(\theta)} + \mathcal{L}^{(\phi)} + \mathcal{L}^{(r)}$,
and the final training objective is represented as
\begin{equation}
\mathcal{L}
=
\mathcal{L}_{\mathrm{traj}}
+ \lambda \mathcal{L}_{\mathrm{beam}},
\end{equation}
where $\lambda$ balances trajectory prediction and beam selection performance, and is empirically set to $10$ in our simulations.

\section{Simulation Results} 

The time-varying near-field channel dataset is generated using the NVIDIA Sionna simulation platform \cite{dorner2022sionna} , following the 3GPP-compliant channel modeling framework. The carrier frequency is set to $f_c = 7$~GHz. The BS is equipped with an XL-MIMO UPA consisting of $M_y \times M_z = 64 \times 64$ antennas. We assume $\sigma^2=1$ for simplicity. 
The generated dataset is divided into three subsets for training (12000 samples), validation (1500 samples), and testing (1500 samples), respectively. Each data sample contains a sequence of $20$ consecutive time slots with a fixed sampling interval of $0.1$s.
The RGB images are resized to $H \times W = 224 \times 224$, while each LiDAR frame consists of $P = 1024$ three-dimensional points. 
For near-field beamforming, a three-dimensional beam codebook is constructed by uniformly sampling the azimuth, elevation, and distance domains, resulting in a codebook size of $N \times N \times S = 20 \times 20 \times 10 = 4000$ candidate beams.

\begin{figure}[!t]
\centering
\includegraphics[height=3.6cm,width=5.2cm]{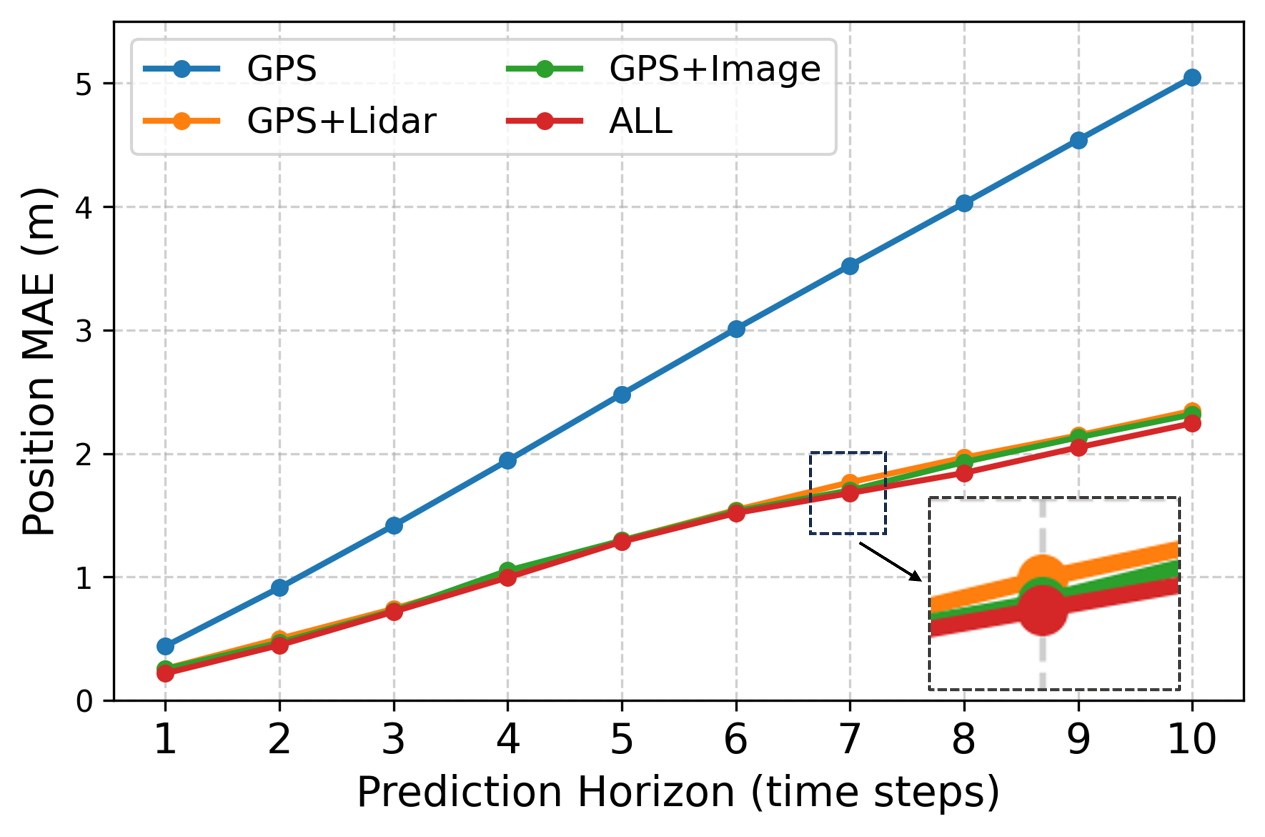}
\caption{MAE performance of different combinations of environmental semantics.}
\label{mae_ablation}
\end{figure}

\begin{figure}[!t]
\centering
\includegraphics[height=3.6cm,width=7.6cm]{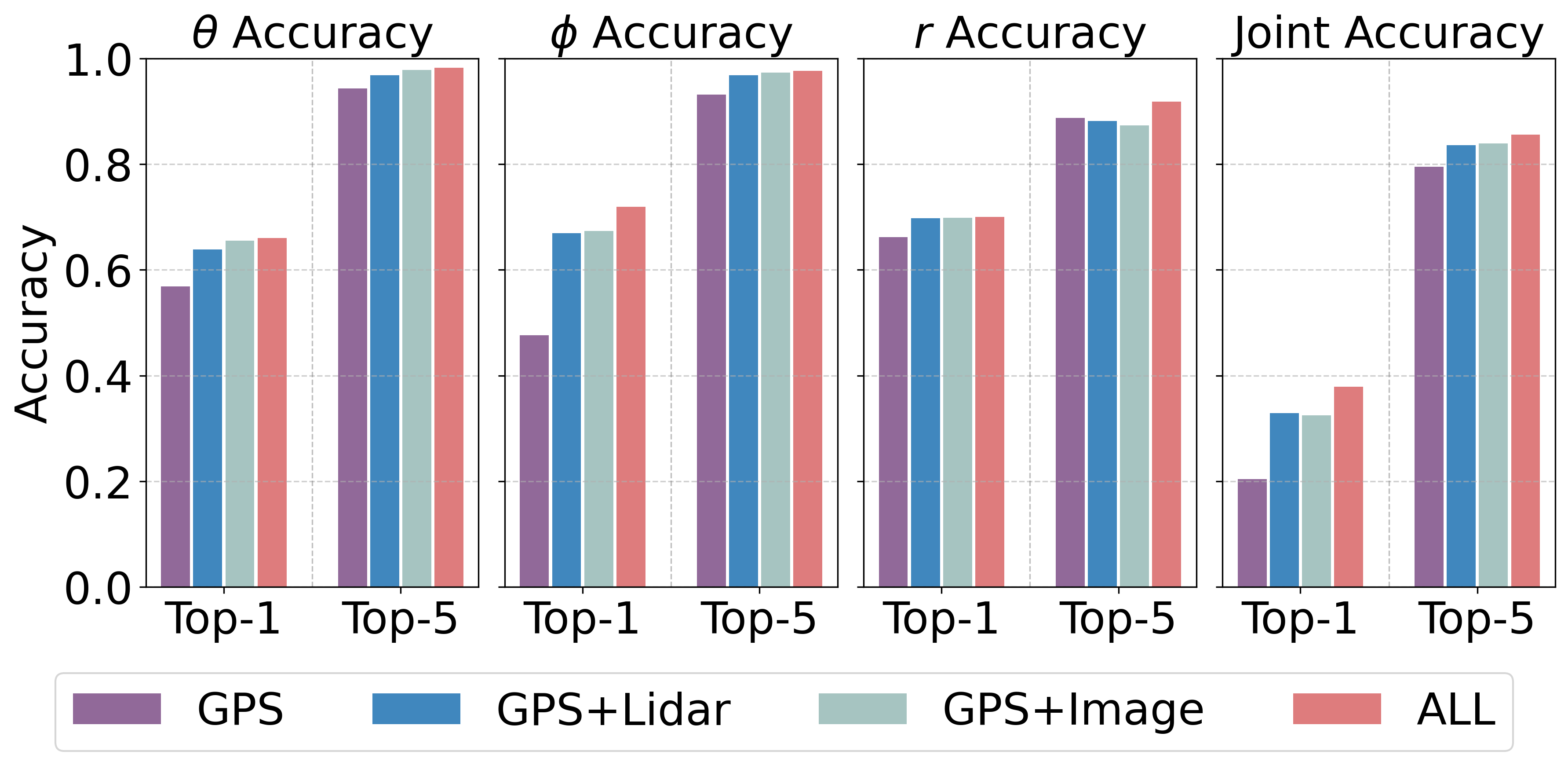}
\caption{Top-$K$ accuracy performance of different combinations of environmental semantics.}
\label{acc_ablation}
\end{figure}

As for performance baselines, we compare the proposed NF-TrackLLM framework with representative sequence learning models, including recurrent neural networks (RNN) \cite{RNN}, long short-term memory (LSTM) \cite{LSTM}, and gated recurrent units (GRU) \cite{1-3}.
For performance evaluation, MAE is used for user localization, while Top-$K$ accuracy is adopted for beam prediction.
Top-1 accuracy indicates whether the optimal beam is correctly predicted, while Top-5 accuracy measures whether it is included among the five highest-ranked candidates.

To investigate the impact of different modalities on beam prediction and user positioning, we conduct an ablation study. 
Fig.~\ref{mae_ablation} shows that the GPS-only scheme suffers from rapid MAE growth with increasing prediction time steps, whereas incorporating LiDAR or image information effectively reduces error accumulation. 
The relatively small gaps among multimodal combinations indicate that environmental semantics mainly act as a stabilizing prior rather than altering the dominant GPS-driven motion trend.
Fig.~\ref{acc_ablation} further demonstrates that multimodal schemes outperform the GPS-only baseline in both Top-1 and Top-5 beam accuracy for $\theta$, $\phi$, $r$, and their joint prediction, with the “ALL” configuration achieving the best overall performance, indicating the effectiveness of environmental semantics for both trajectory and beam prediction.

\begin{figure}[!t]
\centering
\includegraphics[height=3.9cm,width=5.6cm]{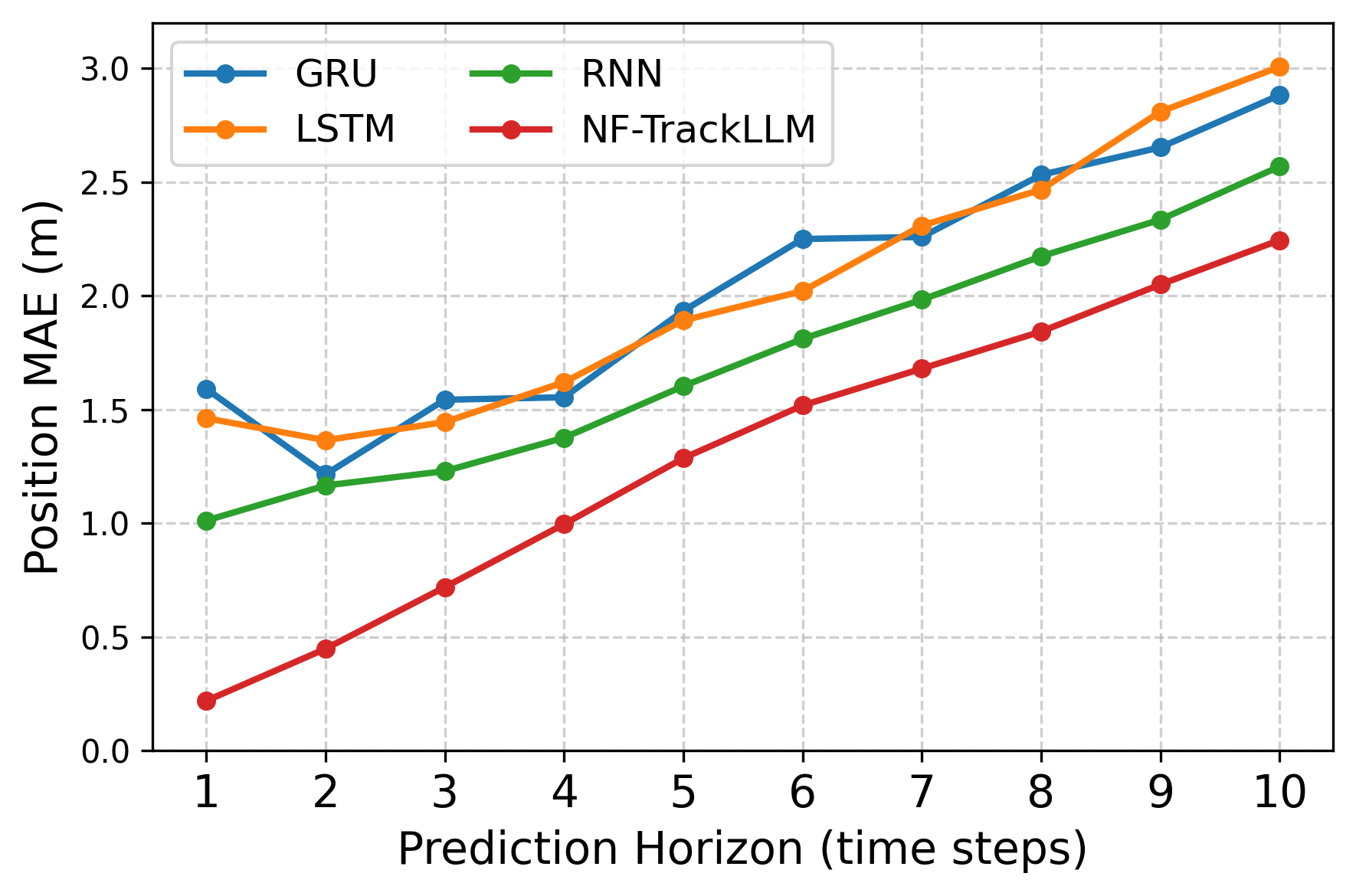}
\caption{MAE performance of different methods.}
\label{mae_baseline}
\end{figure}

\begin{figure}[!t]
\centering
\includegraphics[height=3.6cm,width=7.6cm]{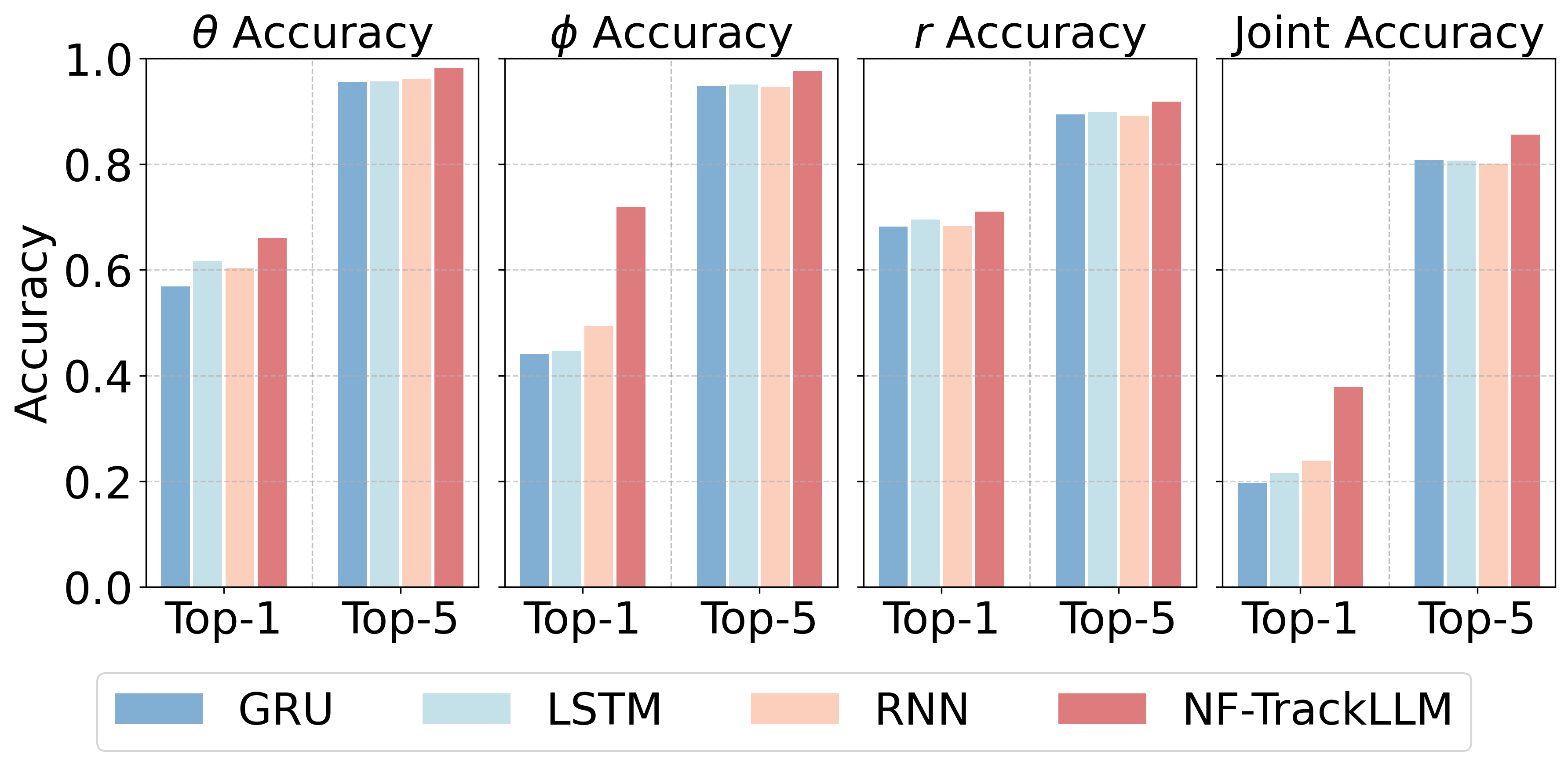}
\caption{Top-$K$ accuracy performance of different methods.}
\label{acc_baseline}
\end{figure}

We further compare the proposed NF-TrackLLM with baselines. As shown in Fig.~\ref{mae_baseline}, NF-TrackLLM consistently achieves the lowest position MAE among all baselines, with an increasingly pronounced advantage at longer prediction time steps. In addition, Fig.~\ref{acc_baseline} indicates that NF-TrackLLM outperforms RNN, GRU, and LSTM in both Top-1 and Top-5 beam accuracy for all components. In particular, it achieves approximately 5\% higher Top-5 joint beam accuracy than the competing methods, demonstrating the superior long-term temporal modeling capability of the proposed NF-TrackLLM.

\begin{figure}[!t]
\centering
\includegraphics[height=4cm,width=7.8cm]{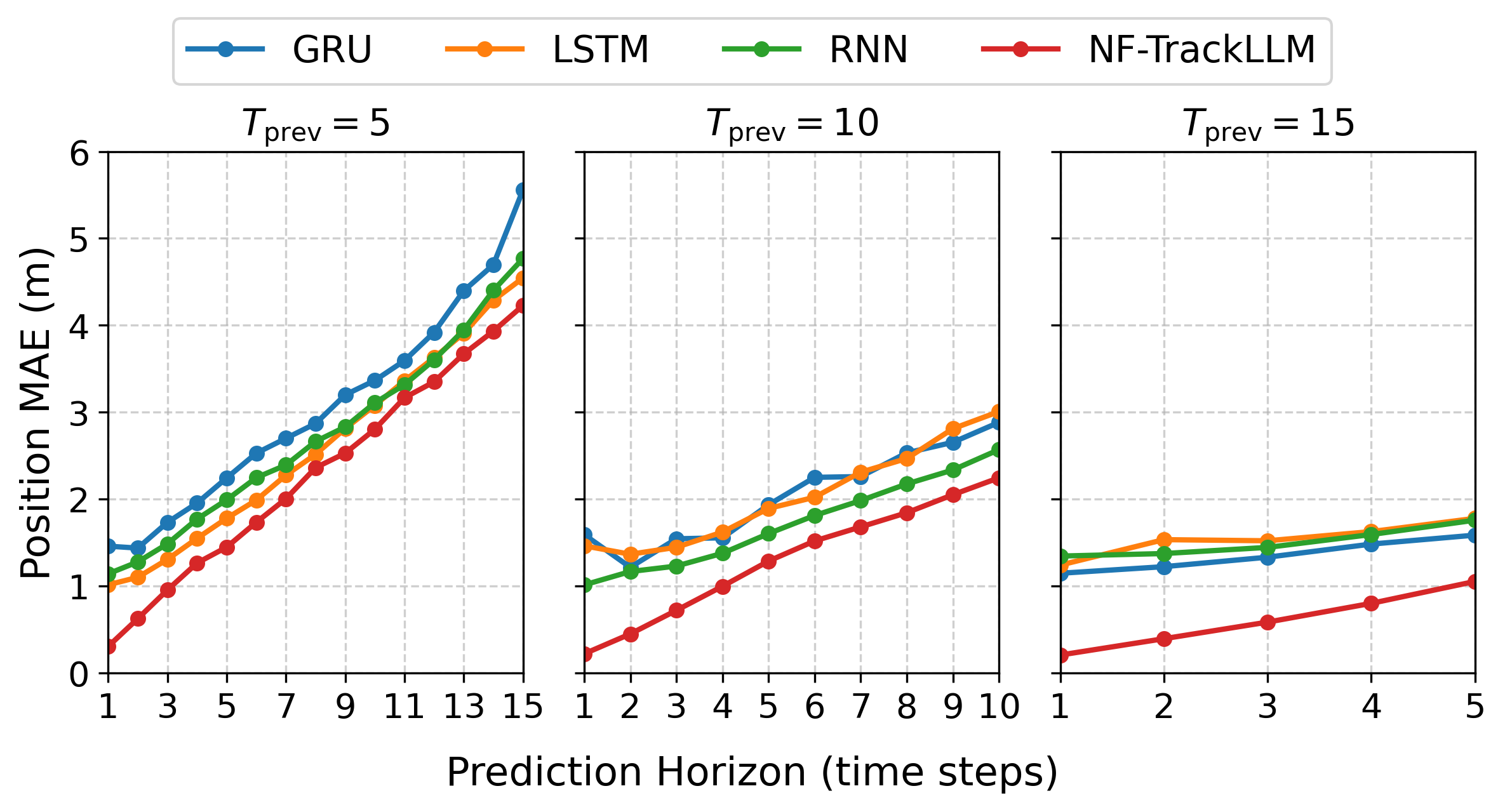}
\caption{MAE performance of different $T_{\mathrm{prev}}$.}
\label{mae_len}
\end{figure}

\begin{figure}[!t]
\centering
\includegraphics[height=4.1cm,width=8.5cm]{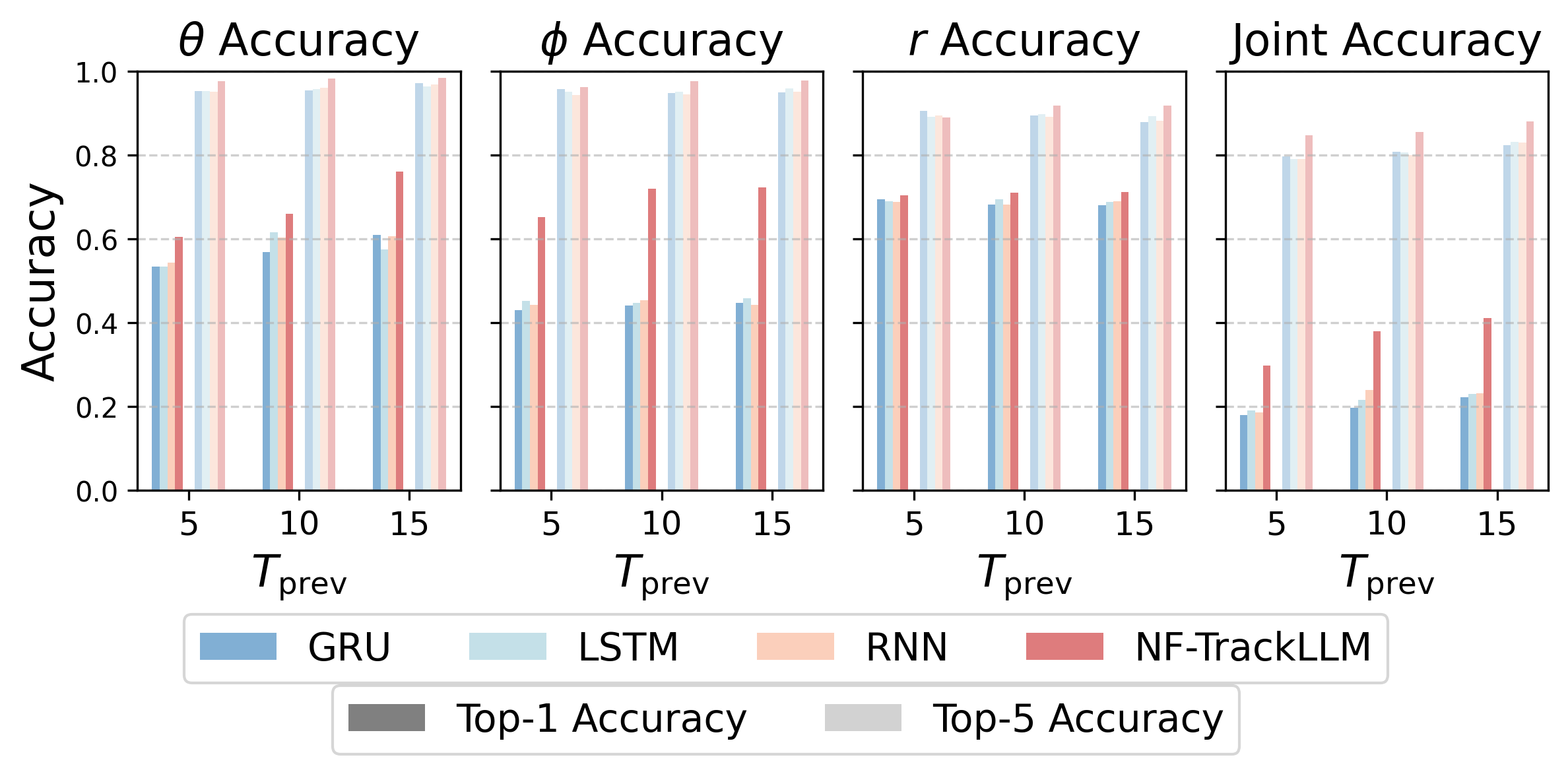}
\caption{Top-$K$ accuracy performance of different $T_{\mathrm{prev}}$.}
\label{acc_len}
\end{figure}

Fig.~\ref{mae_len} and Fig.~\ref{acc_len} show the impact of the historical window length $T_{\mathrm{prev}}$ on trajectory and beam prediction. As $T_{\mathrm{prev}}$ increases from 5 to 15, all methods benefit from richer temporal context, resulting in lower MAE and higher beam accuracy. Meanwhile, NF-TrackLLM achieves the best performance across all window lengths, showing its robustness.

\section{Conclusion}
In this paper, we presented a novel NF-TrackLLM framework for near-field beam prediction and UAV positioning in XL-MIMO systems. By leveraging environmental sensing information and UAV locations through LLM-based semantic reasoning, NF-TrackLLM enables accurate cascaded prediction of UAV trajectories and near-field beams. Simulation results demonstrate that the proposed framework achieves high beam prediction accuracy and reliable trajectory tracking across dynamic urban scenarios, highlighting its potential for practical beam management in future 6G wireless networks.

\section{ACKNOWLEDGEMENT}
This work was supported in part by the National Natural Science Foundation of China (NSFC) under Grants U25A20392, 62422105 and 624B2038.

 \small\bibliographystyle{./bibliography/IEEEtran}
 \bibliography{ref_short}

\vspace{12pt}
\color{red}

\end{document}